# The acceleration and storage of radioactive ions for a neutrino factory

B. Autin, M. Benedikt, S. Hancock, H. Haseroth, A. Jansson\*, U. Köster, M. Lindroos, S. Russenschuck, F. Wenander

*CERN, CH-1211 Geneva 23, Switzerland*

M. Grieser

*Max Planck Institut fuer Kernphysik, 69029 Heidelberg, Germany*

\*) at present: Fermi lab, Batavia, IL 60510, US

## Abstract

The term beta-beam has been coined for the production of a pure beam of electron neutrinos or their antiparticles through the decay of radioactive ions circulating in a storage ring. This concept requires radioactive ions to be accelerated to a Lorentz gamma of 150 for $^6$He and 60 for $^{18}$Ne. The neutrino source itself consists of a storage ring for this energy range, with long straight sections in line with the experiment(s). Such a decay ring does not exist at CERN today, nor does a high-intensity proton source for the production of the radioactive ions. Nevertheless, the existing CERN accelerator infrastructure could be used as this would still represent an important saving for a beta-beam facility. This paper outlines the first study, while some of the more speculative ideas will need further investigations.

## *Introduction*

The evolution of neutrino physics demands new schemes to produce intense, collimated, pure neutrino beams. In the current paper, we discuss the feasibility of a new concept [zuc02] for the production of a single flavour (electron) neutrino beam with a well-known energy spectrum. If combined with an intense pion source for the production of muon neutrinos, the beta-beam can address similar physics issues as the muon neutrino factory [mez02]. The scheme relies on existing technology.

The acceleration of an intense radioactive ion beam to high energies is a new domain in the field of accelerator physics [aut02]. In the following we have limited ourselves to the possibility of basing a facility on parts of the existing CERN infrastructure, namely the Proton Synchrotron (PS) and the Super Proton Synchrotron (SPS).

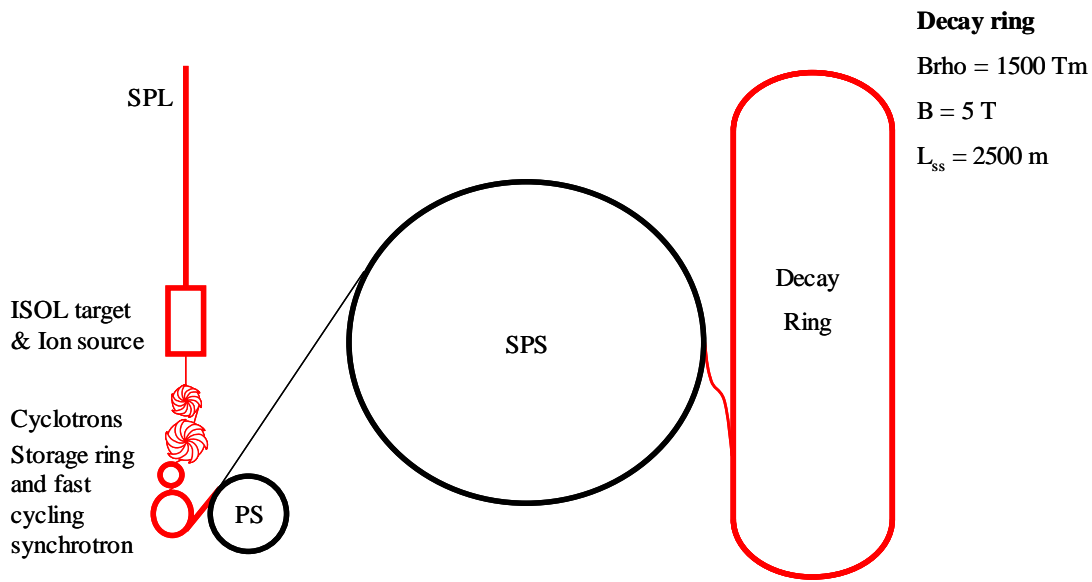

Figure 1: The CERN baseline scenario

## *Baseline scenario*

### Radioactive ion production

The beta-emitting radioactive ions will be produced in an isotope separator on-line (ISOL) system using the proposed Superconducting Proton Linac [spl00] (SPL) as a driver. Plans are being drawn up for a new European radioactive beam facility, EURISOL [eur00], where the method would be exploited to its fullest.

### Ionisation and bunching

The ISOL method produces intense dc beams of ions in low charge states. This beam is accelerated to 50 MeV/u by cyclotrons, injected into a storage ring using charge exchange injection combined with phase space painting, bunched and ejected towards a fast cycling synchrotron. In the latter machine the ions are accelerated to 300 MeV/u and transferred to the PS.

### Acceleration

The PS accumulates 16 bunches one at a time. They are then accelerated, merged in pairs to 8 bunches and transferred to the SPS. The transfer of ions from the PS to the SPS is a well-known space charge bottleneck. In our baseline scenario, bunches fill the maximum available transverse aperture of the SPS and the individual bunch intensity is kept low. The bunches are accelerated in the SPS to the required energy for the chosen ion type. The shortest possible magnetic cycle of the SPS will be used, but it will still induce a dead-time in the production and accumulation which, for SPS top energy, will be close to 8 seconds.

### Transfer to decay ring

The ions are injected in 2 batches of 4 bunches per SPS pulse onto a dispersion-matched orbit in the decay ring and rotated in longitudinal phase space to bring them to the energy of the 4 stored bunches.

### Accumulation in the decay ring

A new merging technique is used to combine the injected bunches longitudinally with those already circulating in the decay ring. The 2 batches are merged one at a time with minimal longitudinal emittance dilution. Many shots can be accumulated until an equilibrium is reached due to beam losses.

## *Production methods for $\nu_e$ emitters*

To exploit a maximum part of the existing CERN accelerator infrastructure requires isotopes that are not too short-lived. For half-lives far below 1 s the decay losses during the acceleration process would become excessive. On the other hand, the half-life should not be too long in order to provide a sufficient production rate in the decay ring.

### $\beta^-$ emitters

Table 1 shows candidate $\beta^-$ emitters. Assuming a limited space charge capacity of the storage ring and completely stripped ions, it is evident that more low-Z isotopes can be stored at a time than high-Z ones. Thus the figure of merit (number of decays per second divided by the average neutrino energy which determines the opening angle of the neutrino beam [zuc02]) is highest for low-Z isotopes. $^8$He and $^9$Li are considered to be too short-lived for efficient acceleration. Thus $^6$He is the best candidate.

| Isotope | A/Z | T ½ (s) | $Q_\beta$ g.s to g.s (MeV) | $Q_\beta$ eff (MeV) | $E_\beta$ av (MeV) | $E_\nu$ av (MeV) | Ions/bunch | Decay rate ($s^{-1}$) | rate / $E_{\nu\ av}$ ($s^{-1}$) |
|---|---|---|---|---|---|---|---|---|---|
| $^6$He | 3.0 | 0.80 | 3.5 | 3.5 | 1.57 | 1.94 | $5 \cdot 10^{12}$ | $4 \cdot 10^{10}$ | $2 \cdot 10^{10}$ |
| $^8$He | 4.0 | 0.11 | 10.7 | 9.1 | 4.35 | 4.80 | $5 \cdot 10^{12}$ | $3 \cdot 10^{11}$ | $6 \cdot 10^{10}$ |
| $^8$Li | 2.7 | 0.83 | 16.0 | 13.0 | 6.24 | 6.72 | $3 \cdot 10^{12}$ | $3 \cdot 10^{11}$ | $4 \cdot 10^9$ |
| $^9$Li | 3.0 | 0.17 | 13.6 | 11.9 | 5.73 | 6.20 | $3 \cdot 10^{12}$ | $1 \cdot 10^{11}$ | $2 \cdot 10^{10}$ |
| $^{11}$Be | 2.8 | 13.8 | 11.5 | 9.8 | 4.65 | 5.11 | $3 \cdot 10^{12}$ | $1 \cdot 10^9$ | $2 \cdot 10^8$ |
| $^{15}$C | 2.5 | 2.44 | 9.8 | 6.4 | 2.87 | 3.55 | $2 \cdot 10^{12}$ | $5 \cdot 10^9$ | $1 \cdot 10^9$ |
| $^{16}$C | 2.7 | 0.74 | 8.0 | 4.5 | 2.05 | 2.46 | $2 \cdot 10^{12}$ | $2 \cdot 10^{10}$ | $6 \cdot 10^9$ |

| Isotope | A/Z | T ½ (s) | $Q_\beta$ g.s. to g.s. (MeV) | $Q_\beta$ eff (MeV) | $E_\beta$ av (MeV) | $E_\nu$ av (MeV) | Ions/bunch | Decay rate (s$^{-1}$) | rate / $E_{\nu\ av}$ (s$^{-1}$) |
|---|---|---|---|---|---|---|---|---|---|
| $^{16}$N | 2.3 | 7.13 | 10.4 | 5.9 | 4.59 | 1.33 | 1·10$^{12}$ | 1·10$^9$ | 1·10$^9$ |
| $^{17}$N | 2.4 | 4.17 | 8.7 | 3.8 | 1.71 | 2.10 | 1·10$^{12}$ | 2·10$^9$ | 1·10$^9$ |
| $^{18}$N | 2.6 | 0.64 | 13.9 | 8.0 | 5.33 | 2.67 | 1·10$^{12}$ | 2·10$^{10}$ | 6·10$^9$ |
| $^{23}$Ne | 2.3 | 37.2 | 4.4 | 4.2 | 1.90 | 2.31 | 1·10$^{12}$ | 2·10$^8$ | 8·10$^7$ |
| $^{25}$Ne | 2.5 | 0.60 | 7.3 | 6.9 | 3.18 | 3.73 | 1·10$^{12}$ | 1·10$^{10}$ | 3·10$^9$ |
| $^{25}$Ne | 2.3 | 59.1 | 3.8 | 3.4 | 1.51 | 1.90 | 9·10$^{11}$ | 1·10$^8$ | 6·10$^7$ |
| $^{26}$Na | 2.4 | 1.07 | 9.3 | 7.2 | 3.34 | 3.81 | 9·10$^{11}$ | 6·10$^9$ | 2·10$^9$ |

Table 1: Candidate isotopes for β$^-$ emitters (charge/bunch = 1×10$^{13}$, γ=100).

## Production of $^6$He

For the production of $^6$He it is preferable to use a direct reaction with high cross-section and little power dissipation of the primary beam. One could consider the $^6$Li(n,p)$^6$He or the $^9$Be(n,α)$^6$He reactions. The former has an energy threshold of $E_n > 2.7$ MeV, the latter of only $E_n > 0.6$ MeV. The cross-section of $^9$Be(n, α) peaks around 100 mb and remains above 25 mb for neutrons between 1.6 and 15 MeV, while the cross-section of $^6$Li(n,p) reaches only 35 mb at maximum. Moreover, Be is more suitable as an ISOL target since it is far more refractory than Li, in particular when bound as BeO.

The required flux of fast neutrons can be produced externally, e.g. by high-energy proton-induced spallation in a heavy metal "converter" mounted close to the ISOL target [Rav02]. With a 100 µA proton beam at 2.2 GeV kinetic energy some 10$^{13}$ $^6$He atoms per second could be produced in the target. Experience with oxide targets [Koe02] shows that the He release from BeO should be faster than from metallic Be. In addition, the former is more refractory allowing stable operation at high temperatures. For all oxide fibre targets discussed in [Koe02], over 80% of the produced $^6$He is released before its decay. Thus, with a beryllia fibre target, which could be heated to still higher temperatures, the efficient release from a large-volume target should also be feasible.

## β$^+$ emitters

Table 2 shows candidate β$^+$ emitters. Boron can react with many elements typically used in ISOL targets and ion sources (C, N, O, metals) and is therefore barely released. No ISOL beams of boron have been produced up to now. $^{33}$Ar is too short-lived for an efficient acceleration in the present scenario and $^{34}$Ar is also rather short-lived. This leaves $^{18}$Ne as the best candidate. As a noble gas it is inert against reactions with the target and ion source materials and can thus be released efficiently even from a bigger target.

| Isotope | A/Z | T ½ (s) | $Q_\beta$ g.s. to g.s. (MeV) | $Q_\beta$ eff (MeV) | $E_\beta$ av (MeV) | $E_\nu$ av (MeV) | Ions/bunch | Decay rate (s$^{-1}$) | rate / $E_{\nu\ av}$ (s$^{-1}$) |
|---|---|---|---|---|---|---|---|---|---|
| $^8$B | 1.6 | 0.77 | 17.0 | 13.9 | 6.55 | 7.37 | 2·10$^{12}$ | 2·10$^{10}$ | 2·10$^9$ |
| $^{10}$C | 1.7 | 19.3 | 2.6 | 1.9 | 0.81 | 1.08 | 2·10$^{12}$ | 6·10$^8$ | 6·10$^8$ |
| $^{14}$O | 1.8 | 70.6 | 4.1 | 1.8 | 0.78 | 1.05 | 1·10$^{12}$ | 1·10$^8$ | 1·10$^8$ |
| $^{15}$O | 1.9 | 122. | 1.7 | 1.7 | 0.74 | 1.00 | 1·10$^{12}$ | 7·10$^7$ | 7·10$^7$ |
| $^{18}$Ne | 1.8 | 1.67 | 3.3 | 3.0 | 1.50 | 1.52 | 1·10$^{12}$ | 4·10$^9$ | 3·10$^9$ |
| $^{19}$Ne | 1.9 | 17.3 | 2.2 | 2.2 | 0.96 | 1.25 | 1·10$^{12}$ | 4·10$^8$ | 3·10$^8$ |
| $^{21}$Na | 1.9 | 22.4 | 2.5 | 2.5 | 1.10 | 1.41 | $9\cdot10^{11}$ | $3\cdot10^{8}$ | $2\cdot10^{8}$ |
| $^{33}$Ar | 1.8 | 0.17 | 10.6 | 8.2 | 3.97 | 4.19 | $6\cdot10^{11}$ | $2\cdot10^{10}$ | $5\cdot10^{9}$ |
| $^{24}$Ar | 1.9 | 0.84 | 5.0 | 5.0 | 2.29 | 2.67 | $6\cdot10^{11}$ | $5\cdot10^{9}$ | $2\cdot10^{9}$ |
| $^{35}$Ar | 1.9 | 1.77 | 4.9 | 4.9 | 2.27 | 2.65 | $6\cdot10^{11}$ | $2\cdot10^{9}$ | $8\cdot10^{8}$ |
| $^{37}$K | 1.9 | 1.22 | 5.1 | 5.1 | 2.35 | 2.72 | $5\cdot10^{11}$ | $3\cdot10^{9}$ | $1\cdot10^{9}$ |
| $^{80}$Rb | 2.2 | 34 | 4.7 | 4.5 | 2.04 | 2.48 | $3\cdot10^{11}$ | $6\cdot10^{7}$ | $2\cdot10^{7}$ |

Table 2. Candidate isotopes for $\beta^+$ emitters (charges/bunch = $1\times10^{13}$, $\gamma=100$).

## Production of $^{18}$Ne

$^{18}$Ne can be produced by spallation in a target (Na, Mg, Al, Si) with cross-sections of the order of 1 mb at 2.2 GeV. Candidate compounds for an ISOL target are e.g. MgO, MgS, $Al_2O_3$, $Al_4C_3$ or SiC. Using, for example, a 1 m long MgO target of 20% theoretical density would produce about $1\times10^{10}$ $^{18}$Ne per µC of primary proton beam, i.e. $1\times10^{12}$ $^{18}$Ne per second from a 100 µA proton beam. Note that the 2.2 GeV protons lose only about 130 MeV of their energy when traversing such a target. Thus, in principle, the exiting proton beam could be sent onto a secondary production target behind. To avoid a local overheating of the target material, the proton beam has to be spread or scanned over a sufficiently large target cross-section to disperse the 13 kW beam power. Thus the target volume will reach several $dm^3$. It still needs to be studied how efficient the release of $^{18}$Ne from such a target will be. To handle a further increase of the proton beam intensity would require a still bigger target or a system of multiple independent targets.

### *Ionisation scenarios for $\nu_e$ emitters*

The produced radioactive elements effuse out of the target container as neutral atoms, and must efficiently and rapidly be ionised in an ion source to reduce the decay losses. The radioactive gas flux from the target is semi-continuous, as the driver beam repetition rate is 50 or 75 Hz. Thus, ideally, the radioactive elements should be collected and ionised during the ramping time of the SPS and then extracted with a pulse length of <100 µs for fast injection into the circular machines. With present technology, this is several orders of magnitude away in terms of the required space-charge capacity. Nevertheless, alternative solutions may be viable. Firstly, the space charge can be reduced by shortening the collection time to ~2.5 half-lives, that is to 2 s for $^6$He and 4 s for $^{18}$Ne, as the particle gain for longer collection times is negligible due to decay. Secondly, highly efficient ion source concepts have recently been developed.

### ECR ion source alternative

A compact ECR ion source, with high ionisation efficiency for noble gases (45% for He [Jar02a] and >90% for Ne [Jar02a,Oya98]), connected directly to the target outlet minimises the effusion delay time. The extracted dc beam consists mainly of $He^+$ ($Ne^+$), with a $He^{2+}$ ($Ne^{n+}$) fraction of a few percent. The ionisation time of 50 and 150 ms for 90% of the total number of ions for He and Ne [Jar02b] is relatively short compared with the half-life. Such a source has no ion-storing capability and the ions leave the plasma volume continuously within some milliseconds after ionisation[1]. Assuming the radioactive gas fluxes above, a radioactive current of 10 µA is reached. This is far below the several mA that an ECR source is capable of delivering. A cold transfer line between target and ion source suppresses the influx of condensable elements. A separator magnet is inserted after the source to separate the radioactive ions from the carrier gas ions.

---

[1] An electrostatic pulsed extraction has been shown to have a limited blocking effect on the beam and works efficiently mainly for repetition frequencies above 500 Hz [Jeo96].

## *Injection into the storage ring*

The requirements for the transverse emittance in the decay ring for the beta-beam are relaxed. This helps to overcome the space charge bottleneck between the PS and SPS. It also sets a generous upper limit for the physical emittance in the storage ring (see Table 3), which is important considering the long injection times required to transform the dc beam from the cyclotrons to a bunched beam suitable for synchrotrons. The beam will be injected using a combination of charge-exchange injection through a thin foil and phase space painting. The latter will reduce the number of passages through the foil for each ion, which will reduce losses and angular straggling of the ions. This process has not been studied in detail but charge exchange injection for stable He and Ne ions is routinely used at the The Svedberg laboratory in Uppsala [rei00].

| Machine | | Kinetic energy | Physical emittance $\pi$ mm mrad | Normalised emittance $\pi$ mm mrad |
|---|---|---|---|---|
| ECR | | 20 keV/u | 50 | 0.5 |
| Cyclotron | | 50 MeV/u | 1.5 | 0.5 |
| Storage Ring | | 50 MeV/u | 78 | 26 |
| Fast Cycling Syncrotron | | 300 MeV/u | 30 | 26 |
| PS | He | 3.5 GeV/u | 20 | 93 |
| | Ne | 7.8 GeV/u | 20 | 186 |
| SPS | He | 139 GeV/u | 0.6 | 93 |
| | Ne | 55 GeV/u | 3.1 | 186 |
| Decay Ring | He | 139 GeV/u | 0.6 | 93 |
| | Ne | 55 GeV/u | 3.1 | 186 |

Table 3: Transverse vertical emittance of the beam ejected from each machine. The limitation in the horizontal plane is less severe in the existing CERN machines. The normalized emittance is increased by phase space painting during charge exchange injection in the storage ring. A blow-up foil is used in the PS and SPS to reduce space charge effects in the following machines.

## *Space Charge Bottleneck at SPS Injection*

Taking the so-called "ultimate" LHC proton beam at 26 GeV/c ($\Delta Q_V = -0.07$) to benchmark the space charge limit at SPS injection leads to the figures given in Table 4.

| | $N_b^{Max}$ | $N_b^{Baseline}$ | Missing Factor |
|---|---|---|---|
| p | $1.7 \times 10^{11}$ | | |
| $^6He^{2+}$ | $9.4 \times 10^9$ | $1.2 \times 10^{12}$ | 130 |
| $^{18}Ne^{10+}$ | $5.2 \times 10^9$ | $6.5 \times 10^{10}$ | 13 |

Table 4: Space charge limits, $N_b^{Max}$, at SPS injection as given by the ultimate LHC beam.

The SPS was designed for fixed-target physics. The machine is well adapted to handle beams with small momentum spread, moderate bunch intensity and large transverse emittance. The LHC beam has large momentum spread, high bunch intensity and small transverse emittance. In fact, the physical emittance is only of the order of 1 µm at SPS injection, whereas the vertical acceptance approaches 20 µm. This alone should allow the missing factor in Table 4 to be reduced by more than an order of magnitude. The SPS cycle for the LHC involves a long wait for up to 4 PS batches, whereas the single beta-beam bunch could even be injected into a moving bucket. This means that the beta-beam could probably tolerate a larger initial tune shift. A further factor of 5 could be gained by installing a moderate (~1 MV) 40 MHz rf system in the SPS. This would be sufficient to accelerate the ions to near transition, where the bunch would naturally be short enough for the standard 200 MHz system to take over.

### *Induced radiation in the machines*

Since the radioactive nuclei have a relatively short lifetime, a large portion of the initial beam will decay during acceleration. Activation of the machines will therefore be an issue. As nuclei change their charge in beta-decay, one could imagine a design for the new purpose-built low-energy machines such that most of these decays occur in the straight sections and the magnets act as separators directing the decay products to dedicated beam dumps. In the existing machines, this might not be so easy. In the PS, for example, there are no long straight sections, so the decay losses would be more or less evenly distributed in the machine.

The total deposited power can be written

$$P = N \frac{E_0 \ln 2}{t_{1/2}} \frac{\gamma - 1}{\gamma} ,$$

where $N$ is the number of particles, $E_0$ is the rest energy of the nuclei, and $t_{1/2}$ is the half-life at rest of the ion species. The factor $\gamma$-1 comes from the kinetic energy, and $\gamma^{-1}$ from the time dilatation. For sufficiently high values of $\gamma$, the loss power is thus energy independent; it only depends on the number of particles in the machine.

Averaging over the acceleration cycle and assuming that losses are evenly distributed around the machine, one obtains the power per unit length. This is what ultimately determines the activation of the machine. Typically, 1 W/m is quoted as an acceptable upper limit, since for 1 GeV protons it produces an activation just below the US limit for "hands-on" maintenance (100 rem). However, the activation is energy dependent. Simulations made for the SNS show that the activation for a fixed loss power increases with energy up to 1 GeV [har99]. Analytic calculations show that, since high-energy particles are not absorbed in the machine components, the machine activation actually decreases with energy at higher energies [sul92]. Instead, the particles traverse the machine and activate the shielding.

The average deposited power due to beta decay, calculated for the PS and SPS, is given in Table 5. It can be seen that the PS is just above the 1 W/m limit. Of course, one must also add normal losses to these numbers.

| Machine | Ions extracted | Batches | Loss power | Losses/length |
|---|---|---|---|---|
| Source + Cyclotron | $2\ 10^{13}$ ions/s | 52.5 ms | N/A | N/A |
| Storage Ring | $1.02\ 10^{12}$ | 1 | 2.95 W | 19 mW/m |
| Fast Cycling Synchrotron | $1.00\ 10^{12}$ | 16 | 7.42 W | 47 mW/m |
| PS | $1.01\ 10^{13}$ | 1 | 765 W | 1.2 W/m |
| SPS | $0.95\ 10^{13}$ | ∞ | 3.63 kW | 0.41 W/m |
| Decay Ring | $2.02\ 10^{14}$ | N/A | 157 kW | 8.9 W/m |

| Machine | Ions extracted | Batches | Loss power | Losses/length |
|---|---|---|---|---|
| Source + Cyclotron | $8\ 10^{11}$ ions/s | 52.5 ms | N/A | N/A |
| Storage Ring | $4.14\ 10^{10}$ | 1 | 0.18 W | 1.1 mW/m |
| Fast Cycling Synchrotron | $4.09\ 10^{10}$ | 16 | 0.46 W | 2.9 mW/m |
| PS | $5.19\ 10^{11}$ | 1 | 56.4 W | 90 mW/m |
| SPS | $4.90\ 10^{11}$ | ∞ | 277 W | 32 mW/m |
| Decay Ring | $9.11\ 10^{12}$ | N/A | 10.6 kW | 0.6 W/m |

Table 5: Intensities and average loss power for the $^6$He (top) and $^{18}$Ne (bottom) beam, assuming a 16 Hz fast cycling synchrotron and 8 s SPS cycle time. Only beta-decay losses are taken into account.

### *Losses in the decay ring*

The losses in the high-energy storage ring can be estimated using the fact that no beam is ejected. All injected beam is essentially lost somewhere in the machine. Hence

$$P_{tot} = \frac{N_{inj} E}{t_{rep}}$$

where $N_{inj}$ in the number ions in each injected batch, $E$ is the kinetic energy of the ions, and $t_{rep}$ is the injection repetition rate.

With the proposed layout of the decay ring, about 14 % of the beta decay products end up in each arc, and 36 % decay in each of the straight sections. For $^6$He, this correspond to an energy deposition of 8.9 W/m in the arcs, and 56 kW in a hot spot downstream of the first bend after the straight sections. To control losses in the straight sections, some kind of separation scheme at the end of each section could be employed to separate decay products from the beam and dump them in a controlled way.

The magnitude of the losses probably excludes the use of superconducting magnets, thereby increasing the length of the decay ring.

The losses that are not due to beta decay will be dominated by longitudinal acceptance limitations due to the stacking method. Momentum collimation might be required to control these losses.

## Cooling

To reach the desired intensity, stacking will be required in the high energy decay ring. Without cooling, Liouville's theorem restricts the stacking process. If electron cooling could be used in the decay ring, it could increase the stacking efficiency. All operational electron coolers today work below about 1 GeV/u. High and medium energy electron cooling is currently investigated at Brookhaven [bur00], Fermilab [nag00]and DESY [bal00]. Calculations carried out for 150 GeV protons in the Tevatron yield cooling times of about 5 minutes [der00]. Re-scaling this result for $^6$He$^{2+}$ and $^{18}$Ne$^{10+}$ gives 7.5 and 1.4 minutes, respectively. Given that the injection repetition rate in the decay ring is 8 s, electron cooling would therefore not have any significant effect on the stacking efficiency. Stochastic cooling is also excluded. This is because there are too many particles per bunch.

## Bunch rotation stacking in the beta-beam decay ring

The decay ring is an accumulator of the bunches delivered by the injector chain. Accumulation is required because the half-life of the stored ions is more than an order of magnitude longer than the cycling time of the injectors. It is complicated by the need to stack the beam in only a few bunches and by the fact that cooling is excluded. One approach is to use asymmetric bunch pair merging, which combines adjacent bunches in longitudinal phase space such that a small bunch can be embedded in the densest region of a much larger one with minimal emittance dilution.

A fresh bunch must be injected in the neighbouring bucket to an existing bunch in the stack, but this is excluded using conventional kickers and septa because of the short rise time that would be required. An alternative injection scheme exploits the fact that the stack is located at only one azimuth in the decay ring and that the revolution period is relatively long. The new bunches are off momentum and are injected in a high dispersion region on a matched dispersion trajectory. Subsequently, each injected bunch rotates a quarter turn in longitudinal phase space until the initial conditions for bunch pair merging are met.

The starting point is a series of 4 consecutive stack bunches in a dual-harmonic system in the decay ring. In order to satisfy the bunch length requirements imposed by the experiments, 40 and 80 MHz rf systems are needed. The total length of the four-bunch train is approximately 1 µs, thus occupying 1/20 of the circumference of the machine.

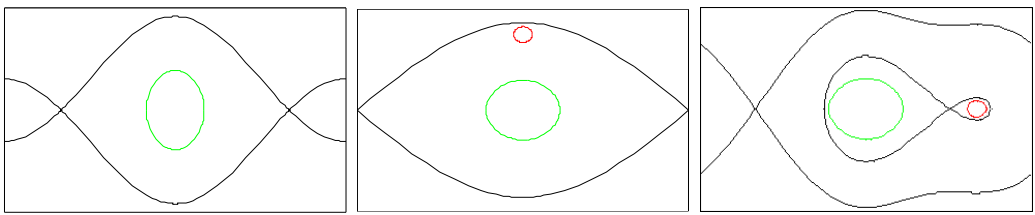

Figure 2: Bunch rotation stacking, longitudinal phase space plots (Energy versa Time): Left, Steady-state stacked bunch in decay mode. Middle, Injected and stacked bunches. Right, Start of bunch pair merging. The horizontal time axis on each plot is 25 ns.

Prior to injection, the second harmonic component is reduced to zero and local closed-orbit bump pushes the circulating bunches towards the blade of the magnetic injection septum. Each new bunch arrives in phase with a circulating one, but separated from it in momentum by an amount which provides the space for the septum blade in the dispersion region.

The local orbit bump must collapse sufficiently during one turn (~20 μs) to bring the injected bunches across the septum blade. One quarter of a synchrotron period after injection, each new bunch has rotated to the same momentum as the stacked ones and a suitably phased second-harmonic component is snapped on. This is the starting point for asymmetric bunch pair merging.

Given that the half-lives of $^6$He and $^{18}$Ne are both of the order of two minutes at their respective top energies while the cycling time of the injector chain is of the order of 8 s, it is clear that each bunch of the stack will have a longitudinal emittance that is more than an order of magnitude larger than that of an incoming bunch. Asymmetric bunch pair merging allows the fresh, dense bunch to be deposited at the centre of the large accumulated one. Thus the oldest ions are moved to the edge of the stack and, due to their decay, a steady state is reached. Bunch characteristics throughout the baseline scenario are presented in Table 6.

| Machine | Number of bunches | | Final bunch length (ns) |
|---|---|---|---|
| | Injection | Ejection | |
| Storage ring | CW beam | 1 | Not evaluated |
| Fast cycling synchrotron | 1 | 1 | Not evaluated |
| PS | 16 | 8 | 20 |
| SPS | 8 | 2 x 4 | 1 |
| Decay ring | 4 | - | <10 |

Table 6: Bunch characteristics.

## Merging simulation

As a proof of principle, the accumulation of a complete stack has been crudely simulated (using the SPS as a model for the decay ring). The full-blown scheme sees two batches each of four bunches transferred and stacked in the decay ring. This takes of the order of one second.

Asymmetric merging is achieved by controlling the relative phase of the two rf components as a function of their decreasing voltage ratio such that the acceptance of the inner bucket containing the stack bunch is gradually reduced while that containing the fresh bunch is maintained. The simulation simply took a single injected bunch of $^6$He after its quarter turn rotation and stacked this particle distribution again and again. At each repetition, some of the resultant stack was removed at random corresponding to the expected number of $^6$He decays. A steady state was reached at an intensity, which was within 20% of that which would have been achieved with a stacking efficiency of 100%. This revealed that, provided the emittance of the injected bunch can be kept below 1 eVs, the order of magnitude of the rf voltages required for merging is restricted to a comparatively modest 10 MV. The final intensity of a bunch in the decay ring can exceed ten times that of an injected bunch.

## *Conclusions*

A possible scenario for accelerating radioactive ions for a beta-beam facility has been developed. It makes use of large parts of the existing CERN accelerator infrastructure and ties up with other CERN activities, such as ISOLDE and the muon neutrino beam. Several possible showstoppers have been circumvented, but much work is still required if the facility is ever to be built. Especially, using the concept of charge exchange injection for the bunching of the DC beam in the low energy accumulator ring needs further analysis. Furthermore, machine activation is a major problem and only a detailed study can show if it will be possible to handle the losses. Still, it is important that the beta-beam concept be studied. A "green field" scenario free from the limitations imposed by the existing CERN accelerator infrastructure should also be considered for a complete picture of the possibilities offered by this exciting concept.